\documentclass{iopart}
\usepackage{iopams}
\usepackage{graphicx}

\renewcommand{\vec}[1]{\mathbf{#1}}
\newcommand{\eqref}[1]{(\ref{#1})}

\makeatletter

\begin{document}

\title{Resonant scattering effect in spectroscopies of interacting atomic gases}

\author{M.J. Leskinen, J. Kajala, and J.J. Kinnunen}
\ead{jami.kinnunen@tkk.fi}
\address{Department of Applied Physics, P.O. Box 5100, 02015 Aalto University, Finland}

\pacs{03.75.Ss, 32.80.-t, 73.50.Lw}

\begin{abstract}
We consider spectroscopies of strongly interacting atomic gases, and
we propose a model for describing the coupling between quasiparticles
and gapless phonon-like modes. Our model explains features
in a wide range of different experiments in both fermionic and bosonic atom 
gases in various spectroscopic methods.
\end{abstract}

\submitto{\NJP}

\maketitle

\section{Introduction}

Experimental setups in ultracold atomic gases allowing easy optical 
access to the sample and the well established imaging techniques 
have resulted in a wide range of precise and well understood 
spectroscopies. Radio-frequency~\cite{GGupta2003a, Regal2003a, CChin2004a, 
Moritz2005a, BBartenstein2005a, SSchunck2007a, SSchunck2008a, Stewart2008a} (RF), Raman~\cite{Blakie2006a,Dao2007a}, Bragg~\cite{Stenger1999a,Veeravalli2008a,Papp2008a,Fabbri2009a}, and lattice modulation spectroscopies~\cite{Stoferle2004a, Schori2004a, Jordens2008a} have plenty in 
common and they can all be described effectively as an absorption of a single
photon of momentum $\vec{k}$ and frequency $\omega$. 

The agreement between spectroscopic experiments and theories in
ultracold gases has often 
been at most qualitative despite the various theoretical approaches 
such as mean-field~\cite{Torma2000a, KKinnunen2004b, Ohashi2005b,He2005a,Yu2006a,Basu2007a,MMassignan2007a,Kinnunen2009a,Chen2009a}, 
t-matrix~\cite{Punk2007a}, or other diagrammatic 
approaches~\cite{Kinnunen2006b,PPerali2008a,HHe2009a,Pieri2009a}.
Strange peaks and features (such as anomalously broad spectral peaks) 
in the experiments have been 
attributed to the lack of coherence~\cite{Fabbri2009a}, defects in a lattice~\cite{Stoferle2004a}, or to 
center-of-mass motion of pairs~\cite{Stewart2008a}, suggesting that
those features are caused by imperfections in the experimental setups.
These anomalies have been all but neglected 
in theoretical descriptions and, instead, a lot of effort has been
put into improving the underlying many-body theories by including thermal and
quantum fluctuations~\cite{Haussmann1994a, Chen2005a, Perali2002a} 
in order to obtain quantitative agreement with the well understood features
in the spectra. Here we show that those neglected anomalies
seen in a wide range of different experiments have a common source.

While the breakdown of a mean-field theory in strongly interacting systems is fully anticipated, the treatment of
beyond mean-field effects is exceedingly difficult. As a result most theoretical studies have relied on the mean-field
approximation, partly because of the apparently good agreement between theory and experiments. However, as we will soon
point out, various experiments performed with strongly interacting systems have features 
that cannot be explained by mean-field theories. Here we propose an efficient diagrammatic expansion for the
mean-field theories to describe an important class of coherent processes.

\section{Linear response spectroscopy}

We begin by deriving the linear response theory for a weak perturbation $V$.
While the calculation is a simple textbook derivation, it will help to 
illuminate the problem encountered with the mean-field linear response theory.

Let's assume that the system is initially (at time $t=0$) in a ground
state of a (many-body) Hamiltonian $H_0$, denoted by $|\psi(0)\rangle$.
Switching on perturbation $V(t)$ yields
the full time-dependent Hamiltonian $H(t) = H_0 + V(t)$, and
the system evolves according to the time-dependent 
Schr\"odinger equation
\[
   i\hbar \frac{d}{dt} |\psi(t)\rangle = H(t) |\psi(t)\rangle.
\]
This can be solved to yield the state at time $t$ as
\[
  |\psi(t)\rangle = e^{-iH_0t/\hbar} S(t) |\psi(0)\rangle,
\]
where the scattering S-matrix is
\[
 S(t) = 1 + \frac{1}{i\hbar} \int_0^t dt'\, V_\mathrm{I} (t') + \left(\frac{1}{i\hbar}\right)^2 \int_0^t dt'\, \int_0^{t'} dt''\, V_\mathrm{I} (t') V_\mathrm{I} (t'') + \ldots,
\]
where $V_\mathrm{I}(t) = e^{iH_0t/\hbar} V(t) e^{-iH_0 t/\hbar}$
is the perturbation expressed in the interaction picture.
This formulation of S-matrix readily offers easy access to the
perturbative description. Cutting the series expansion of the S-matrix
yields perturbation theories of different orders. For our purposes,
it is enough to consider the first three terms of the above expansion,
yielding the approximate wavefunction in the second-order perturbation 
theory
\begin{equation}
\fl   |\psi(t)\rangle \approx e^{-iH_0t/\hbar} \left[ 1 + \frac{1}{i\hbar} \int_0^t dt'\, V_\mathrm{I} (t') + \left(\frac{1}{i\hbar}\right)^2 \int_0^t dt'\, \int_0^{t'} dt''\, V_\mathrm{I} (t') V_\mathrm{I} (t'') \right] |\psi(0)\rangle.
\label{eq:wavef}
\end{equation}

Ultimately, we are interested in the expectation value of some 
observable $\mathcal{O}$. Keeping terms only up to the second order in
the perturbation $V$, we obtain
\begin{eqnarray}
\fl  \langle \mathcal{O} (t) \rangle =& \langle \psi (t) | \mathcal{O} | \psi(t) \rangle = \langle \mathcal{O} \rangle +\frac{1}{\hbar^2} \int_0^t dt'\, \int_0^t dt''\, \langle V_\mathrm{I} (t')\mathcal{O}_\mathrm{I}(t) V_\mathrm{I} (t'')\rangle \nonumber \\ 
\fl   &-\frac{1}{\hbar^2} \int_0^t dt'\, \int_0^{t'} dt''\, \left[ \langle \mathcal{O}_\mathrm{I}(t) V_\mathrm{I} (t') V_\mathrm{I} (t'')\rangle + \langle V_\mathrm{I} (t'') V_\mathrm{I} (t') \mathcal{O}_\mathrm{I}(t)\rangle \right],
\label{eq:expvalue}
\end{eqnarray}
where the expectation values on the right hand side are calculated for the initial state $|\psi(0)\rangle$.
Should we be interested in the rate of change of 
$\langle \mathcal{O} (t) \rangle$ (as is the case when measuring current), 
one can differentiate Eq.~\eqref{eq:expvalue} with respect to time $t$ and obtain
\begin{equation}
  \frac{d}{dt} \langle \mathcal{O} (t) \rangle = \frac{2}{\hbar^2} \int_0^t dt'\, \Re \left[ \langle V_\mathrm{I}(t) \mathcal{O}_\mathrm{I} (t) V_\mathrm{I}(t')\rangle - \langle \mathcal{O}_\mathrm{I} (t) V_\mathrm{I}(t) V_\mathrm{I}(t') \rangle \right],
\label{eq:rate}
\end{equation}
where $\Re$ is the real-part and we have assumed here that $V(t)$ and 
$\mathcal{O}$ are Hermitian operators.
In order to continue this discussion on linear response, we need to specify
the perturbation $V$ and the observable $\mathcal{O}$.

\subsection{Radio-frequency spectroscopy}

We will first consider the case of radio-frequency spectroscopy.
A radio-frequency field of frequency $\omega$ and amplitude $\Omega$ drives 
atoms from hyperfine state $|2\rangle$ to \emph{initially unoccupied} 
hyperfine state $|3\rangle$. As the wavelength of the radio-frequency field 
is much
longer than any relevant length scale in the system, the radio-frequency 
photons do not affect the momentum of the atoms but only couple to the 
hyperfine state degrees of freedom.
The perturbation in the rotating wave approximation is then given by
$V(t) = \Omega e^{i\delta t/\hbar} \sum_k c_{3k}^\dagger c_{2k} + H.c.$,
where $\delta$ is the detuning of the rf-field frequency away from the
Zeeman splitting of hyperfine states $|2\rangle$ and $|3\rangle$,
$\delta = \hbar \omega + E_\mathrm{hf}^3 - E_\mathrm{hf}^2$,
where $E_\mathrm{hf}^\sigma$ is the hyperfine energy of state 
$|\sigma\rangle$.
The observable of interest is the number of atoms in hyperfine state 
$|3\rangle$: $\mathcal{O} = N_3 = \sum_k c_{3k}^\dagger c_{3k}$.
Since the hyperfine state $|3\rangle$ is initially empty, 
$N_3 |\psi(0) \rangle = 0$, 
the formula for the current simplifies into the form:
\begin{equation}
  \frac{d}{dt} \langle N_3 (t) \rangle = \frac{2}{\hbar^2} \int_0^t dt'\, \Re \langle e^{iH_0t/\hbar} V(t) N_{3} e^{-iH_0(t-t')/\hbar} V (t') e^{-iH_0t'/\hbar}\rangle,
\label{eq:current0}
\end{equation}
where we have written the operators in the Schr\"odinger picture.
Operating by $N_3$ within the expectation value on the right hand side 
of Eq.~\eqref{eq:current0} will necessarily give $1$ since the single 
perturbation $V(t')$ has transferred only one atom into the hyperfine 
state $|3\rangle$ and we assume that $N_3$ commutes with $H_0$. Thus we have
\[
  \frac{d}{dt} \langle N_3 (t) \rangle = \frac{2|\Omega|^2}{\hbar^2} \sum_{kp} \int_0^t dt'\, \Re e^{-i\delta (t-t')/\hbar} \langle c_{2p}^\dagger c_{3p} e^{-iH_0(t-t')/\hbar} c_{3k}^\dagger c_{2k} \rangle,
\]
where we have assumed that the initial state $|\psi(0)\rangle$ is an
eigenstate of $H_0$ with eigenenergy equal to zero (a nonzero energy would provide
only phase factor).
This can be further simplified to yield
\[
  \frac{d}{dt} \langle N_3 (t) \rangle = \frac{2|\Omega|^2}{\hbar^2} \sum_{kp} \Re \int_0^t dt' \, e^{-i\delta t'/\hbar} \langle c_{2p}^\dagger c_{3p} e^{-iH_0t'/\hbar} c_{3k}^\dagger c_{2k} \rangle.
\]
The expectation value on the right-hand side is in principle a 
two-particle (or particle-hole) Green's function propagating within 
a many-body sea of fermions. However, in the special
case where the hyperfine state $|3\rangle$ is noninteracting 
the two-particle propagator separates into a product of one-particle and 
one-hole propagators
\begin{equation}
  \frac{d}{dt} \langle N_3 (t) \rangle = \frac{2|\Omega|^2}{\hbar^2} \sum_{k} \Re \int_0^t dt'\, e^{-i\delta t'/\hbar} G_2 (k,-t') G_3(k,t'),
\end{equation}
where $-iG_2(k,-t') = \langle c_{2k}^\dagger e^{-iH_0t' /\hbar} c_{2k} \rangle$ is the propagator of a hole in single-particle state $|2,k\rangle$ and 
$iG_3(k,t') = \langle c_{3k} e^{-iH_0t' /\hbar} c_{3k}^\dagger \rangle$ 
is the propagator of an atom in state $|3,k\rangle$.
Being a noninteracting Green's function 
(and since hyperfine state $|3\rangle$ is
initially empty), the particle propagator $G_3(k,t')$ 
is simply $-ie^{-i\epsilon_k t'/\hbar}$, where $\epsilon_k$ is the 
energy of an atom with momentum $k$ and we end up with
\[
  \frac{d}{dt} \langle N_3 (t) \rangle = \frac{2|\Omega|^2}{\hbar^2} \sum_{k} \Re \int_0^t dt'\, e^{-i(\epsilon_k + \delta)t'/\hbar} i G_2(k,-t')
\]
In the limit of a very long pulse $t \rightarrow \infty$ (but keeping
the perturbation weak $t \Omega/\hbar \ll 1$), the integral yields
the Fourier transform of the real-time hole propagator. 
\[
  \frac{d}{dt} \langle N_3 (t) \rangle = \frac{2|\Omega|^2}{\hbar^2} \sum_{k} \Im G_2(k,\epsilon_k + \delta),
  \label{eq:linear}
\]
where $\Im$ is the imaginary part. Thus the radio-frequency spectrum probes 
the \emph{spectral function of the hole}
$A_2(k,\epsilon) = 2\Im G_2(k,\epsilon)$ which is ultimately
determined by the operator $e^{-iH_0t/\hbar}$.

While the simple BCS mean-field treatment has proven to be able to describe
spectroscopies of even strongly interacting gases rather well in
many cases, quantitative agreement between theory and the experiments has 
turned out to be very hard to achieve. Therefore, it is worth looking more 
closely what information has been lost
in the BCS mean-field approximation. In particular, we are looking
for non-BCS processes which are important for the hole propagator 
in order to obtain quantitative agreement with the observed spectra.

\subsection{BCS mean-field approximation}

We want to describe radio-frequency spectroscopy of a two-component 
Fermi gas.
Since the rf-field couples one of the two hyperfine states into a third
excited state, the full system is described by a three-component
Hamiltonian but in which the third state $|3\rangle$ is initially empty. 
The Hamiltonian of the three-component system is
\[
  H_0 = \sum_{k\sigma} \epsilon_k c_{k\sigma}^\dagger c_{k\sigma} + \frac{1}{2} \sum_{kpq\sigma\neq \sigma'} g_{\sigma \sigma'} c_{k,\sigma}^\dagger c_{-k+q,\sigma'}^\dagger c_{-p+q,\sigma'} c_{p,\sigma},
\]
where $\epsilon_k$ is the single-particle kinetic energy, 
$\sigma \in \{1,2,3\}$ and $g_{\sigma \sigma'}$ is the interaction
strength between the atoms in hyperfine states $|\sigma\rangle$ and 
$|\sigma'\rangle$ in contact interaction potential approximation.
The standard approach uses the two-body scattering T-matrix with 
$g_{\sigma \sigma'} = \frac{4\pi\hbar^2 a_{\sigma \sigma'}}{m}\frac{1}{V}$,
where $a_{\sigma \sigma'}$ is the scattering length and $V$ is the volume. 
Initially we will assume that only $g_{12}$ is nonzero but later on we will 
consider also the general case in which all interactions can be nonzero.

The BCS mean-field approximation has proven to be very simple but still 
very effective in describing two-component Fermi gases even in the very 
strongly interacting regime. The approximation amounts to first considering
only opposite momentum scatterings $q=0$ in the Hamiltonian $H_0$ and
then replacing the quartic interaction in $H_0$ by quadratic mean-field 
interaction, resulting in BCS mean-field Hamiltonian
\[
  H_0^\mathrm{BCS} = \sum_{k\sigma} \epsilon_k c_{k\sigma}^\dagger 
c_{k\sigma} + \Delta_{12} \sum_{k} c_{k,1}^\dagger c_{-k,2}^\dagger + H.c.,
\]
where the pairing field $\Delta_{12} = g_{12} \sum_p \langle c_{-p,2} c_{p,1} \rangle$. Here we have neglected the Hartree energy shift but it needs to 
be included when studying for example gases in optical lattices in order to 
obtain correct bound state energies.

The BCS mean-field theory can be solved exactly by diagonalizing the
BCS Hamiltonian using the Bogoliubov transformation. This yields
the Hamiltonian in the quasiparticle basis
\[
   H_0^{\mathrm{BCS}'} = \sum_{k\sigma} E_k \gamma_{k,\sigma}^\dagger \gamma_{k,\sigma} + C,
\]
where $C$ is a constant and the operators $\gamma_{k,\sigma}$ ($\gamma_{k,\sigma}^\dagger$) annihilate (create) quasiparticle excitation of momentum $k$ and
energy $E_k = \sqrt{\left(\epsilon_k-\mu\right)^2 + \Delta_{12}^2}$ with
$\mu$ being the chemical potential of the atoms in hyperfine states $|1\rangle$
and $|2\rangle$, and $\epsilon_k$ is the single-particle excitation energy. 
The quasiparticle 'spin' 
$\sigma \in \{ \uparrow, \downarrow \}$ does not refer to the 
hyperfine level of an atom but simply to the fact that there are two branches
of quasiparticle excitations.
The hole spectral function can be shown to be 
\begin{equation}
  G_2(k,\epsilon) = \frac{u_k^2 n_F(E_k)}{\epsilon - E_k + i\eta} + \frac{v_k^2 n_F(-E_k)}{\epsilon + E_k + i \eta},
\label{eq:bcsspectral}
\end{equation}
where the Bogoliubov coefficients $u_k^2 = \frac{1}{2} (1 + \frac{\epsilon_k-\mu}{E_k} )$ and $v_k^2 = 1-u_k^2$.
We have assumed that the system is in equilibrium and the quasiparticle
distributions are given by the Fermi-Dirac distribution 
$n_F(\epsilon) = 1/(1 + e^{\beta \epsilon})$, $\beta = 1/(k_\mathrm{B} T)$, 
$k_\mathrm{B}$ is the Boltzmann constant, and $T$ is the temperature.
We consider only zero temperature $T=0$ results in this manuscript.

At zero temperature, the BCS hole spectral function~\eqref{eq:bcsspectral} 
consists of a single sharp peak (second peak appears at finite 
temperatures), describing 
a quasiparticle excitation with infinite lifetime. This property of infinite 
lifetime can be 
easily seen from the form of the BCS Hamiltonian $H_0^\mathrm{BCS}$.
Creating a hole in state $|k,2\rangle$ will necessarily render any atom
in state $|-k,1\rangle$ effectively noninteracting, since the Hamiltonian
$H_0^\mathrm{BCS}$ describes only scatterings between atoms with opposite
momenta.

In order to consider quasiparticle lifetime effects, one would need to go 
beyond the BCS approximation and include residual interactions between 
the quasiparticles 
\begin{equation}
  H_0^\mathrm{res} = g_{12} \sum_{kp,q\neq0} c_{k,1}^\dagger c_{-k+q,2}^\dagger c_{-p+q,2} c_{p,1}
\label{eq:resint}
\end{equation}
This approach has been studied in the context of nuclear 
physics~\cite{Soloviev} and high-temperature 
superconductors~\cite{Coffey1990a}. It has also been discussed in
the present context in Ref.~\cite{Haussmann2009} using a fully self-consistent
calculation of the many-body T-matrix. The self-consistent nature allowed
the inclusion of Anderson-Bogoliubov (AB) phonon effects, and these were shown
to result in broadening of spectral features. The theory was, however, highly
involved, and the present theory can be seen as an attempt to include important
quasiparticle lifetime effects in a simpler way, namely using first-order 
perturbation theory.  

While Eq.~\eqref{eq:resint} for residual interactions 
is enough for our purposes, it is very instructive to briefly consider
its presentation in the BCS quasiparticle basis.
Writing the residual interaction $H_0^\mathrm{res} = H_{40} + H_{31} + H_{22}$ 
in the quasiparticle basis yields~\cite{Coffey1990a,Haussmann2009}
\begin{eqnarray}
\fl
  H_\mathrm{40} = g_{12} \sum_{kp,q\neq 0} v_{k+q} v_k u_p u_{p+q} \gamma_{k,\uparrow} \gamma_{-k-q,\downarrow} \gamma_{-p,\downarrow} \gamma_{p+q,\uparrow} + H.c.,
\label{eq:quasiham1}
\end{eqnarray}
\begin{eqnarray}
\fl  H_\mathrm{31} = g_{12} \sum_{kp,q\neq 0,\sigma} \left( v_p v_{p+q} v_k u_{k-q} - u_p u_{p+q} u_k v_{k-q} \right) \gamma_{k,\sigma}^\dagger \gamma_{k-q,\sigma} \gamma_{-p,\downarrow} \gamma_{p+q,\uparrow} + H.c.,
\label{eq:quasiham2}
\end{eqnarray}
and
\begin{eqnarray}
\fl  H_\mathrm{22} = g_{12} \sum_{kp,q\neq 0} \left[\left( u_{q-k} u_k u_p u_{q-p} + v_{q-k} v_k v_p v_{q-p} \right) \gamma_{q-k,\uparrow}^\dagger \gamma_{k,\downarrow}^\dagger \gamma_{p,\downarrow} \gamma_{q-p,\uparrow} \right. \nonumber \\
\left.+ \left(u_k u_{p-q} v_p v_{k-q} + v_k v_{p-q} u_p u_{k-q} \right) \gamma_{k,\uparrow}^\dagger \gamma_{-p,\downarrow}^\dagger \gamma_{q-p,\downarrow} \gamma_{k-q,\uparrow} \right.\nonumber \\
\left.+ u_{k+q} u_{p+q} v_k v_p \gamma_{k+q,\uparrow}^\dagger \gamma_{p,\uparrow}^\dagger \gamma_{p+q,\uparrow} \gamma_{k,\uparrow} \right.\nonumber \\
\left.+ v_{k+q} v_{p+q} u_k u_p \gamma_{k+q,\downarrow}^\dagger \gamma_{p,\downarrow}^\dagger \gamma_{p+q,\downarrow} \gamma_{k,\downarrow}\right].
\label{eq:quasiham3}
\end{eqnarray}
Clearly the number of quasiparticles does not need to be conserved
as a single excitation can be split into several
excitations through scattering and vice versa. Moreover,
since quasiparticles have minimum energy cost of $\Delta$, transitions
to states with different numbers of quasiparticles are gapped. 

However, these quasiparticle interactions provide also coupling to the 
collective Anderson-Bogoliubov (AB) mode which is a phonon-like excitation and 
thus gapless. The description of this mode requires nonperturbative treatment,
for example random phase approximation~\cite{Anderson1958a, Minguzzi2001a}, 
and while the AB phonon arises already from the mean-field theory, the 
actual coupling between the quasiparticle
excitations and these collective modes is provided by the residual
interactions $H_0^\mathrm{res}$~\cite{Haussmann2009}. 
What our theory does is to include approximatively the coupling
between the quasiparticle excitation and the phonon modes. We will later 
consider also settings in which the 
couplings to different numbers of quasiparticle excitations will be 
important and
such processes may actually arise naturally from the perturbative treatment.
Furthermore, we will also consider systems at temperatures above the
critical temperature where the superfluid order parameter vanishes. In such case,
the BCS model yields an ideal finite temperature Fermi gas without pairing correlations,
binding energies, or Anderson-Bogoliubov phonons. In that limit, particle and the quasiparticle
become the same.

\section{Beyond BCS mean-field approximation}

Our goal is an effective theory which can include the effect of interactions
between quasiparticles approximatively and which can be easily incorporated 
into the standard mean-field BCS theory. As our starting point we take the 
residual interactions neglected in the BCS mean-field theory 
$H_0^\mathrm{res}$ and treat these as a perturbation in the hole propagator.
Writing the time-evolution operator $e^{-iH_0t}$ in Eq.~\eqref{eq:wavef}
in terms of the scattering S-matrix as
\[ 
   e^{-iH_0t/\hbar} = e^{-iH_0^\mathrm{BCS} t/\hbar} S_\mathrm{res} (t),
\]
and expanding the S-matrix $S_\mathrm{res} (t) = 1 + \frac{1}{i\hbar} \int_0^t dt'\, H_{0\mathrm{I}}^\mathrm{res} (t') + \ldots$, where
$H_{0\mathrm{I}}^\mathrm{res}(t') = e^{iH_0^\mathrm{BCS} t'} H_0^\mathrm{res} e^{-iH_0^\mathrm{BCS} t'}$, we obtain the first-order correction (in terms of $H_0^\mathrm{res}$)
\begin{eqnarray}
\fl   |\psi(t)_1\rangle = \left(\frac{1}{i\hbar}\right)^2 \int_0^t dt' \int_0^{t'} dt''\, e^{-iH_0^\mathrm{BCS} (t-t')/\hbar} \times \nonumber \\
\left[H_{0}^\mathrm{res} e^{-iH_0^\mathrm{BCS} (t'-t'')/\hbar} V (t'') + V (t') e^{-iH_0^\mathrm{BCS} (t'-t'')/\hbar} H_0^\mathrm{res} \right] \times \nonumber \\
e^{-iH_0^\mathrm{BCS}t''/\hbar} |\psi(0)\rangle 
\label{eq:wavef2}
\end{eqnarray}
Here we have included only the first order term (with respect to 
rf-coupling $V$) from Eq.~\eqref{eq:wavef} since the zero and second order 
terms do not contribute to the current in Eq.~\eqref{eq:current0}. The above equation is a result of
applying perturbation theory to both residual interaction $H_0^\mathrm{res}$
and the rf-field $V(t)$. 

At this stage, we notice the similarity with two-photon processes in quantum optics, such as
the two-photon Bragg scattering or two-photon Raman transitions which also
incorporate correlated couplings with two different fields. In the present case
the two fields are the rf-field and the atoms in the hyperfine state $|1\rangle$. 
The main difference, however, is that in our model the atom-atom
interaction has many different scattering channels, including very long
wavelength (low momentum exchange $q$) transitions. Nonetheless, 
the similarity is encouraging - we are here dealing with physics that
has manifested in other systems as well.

\begin{figure}
  \centering
  \includegraphics[width=0.55\textwidth]{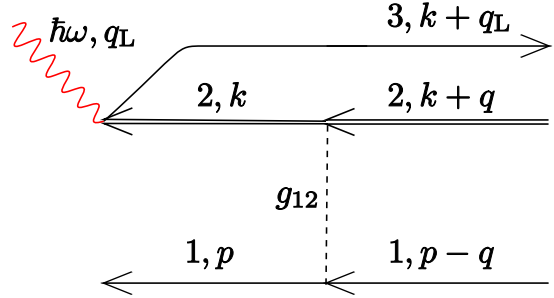}
  \caption{The lowest order photon absorption $\hbar \omega,q_\mathrm{L}$ 
and quasiparticle scattering $g_{12}$ process. The absorption of the photon
by an atom in the ground state of $|2\rangle$ produces 
an atom in hyperfine state $|3\rangle$ and one hole (backward propagator) 
excitation in $|2\rangle$. The hole excitation in $|2,k\rangle$ then 
scatters with a hole in $|1,p\rangle$, resulting in holes at states
$|2,k+q\rangle$ and $|1,p-q\rangle$. As is clear from this description,
the photon absorption and the atom-atom scattering processes are connected
through the hole degrees of freedom.  If the photon has finite 
momentum $q_\mathrm{L}$ such as in Bragg spectroscopy, the atom will also 
feel a momentum change worth $q_\mathrm{L}$. We approximate 
that $|1\rangle$-atoms are described 
by bare propagators and hence the atom-atom scattering 
process does not produce additional excitations experiencing mean-field
energy shifts (such as BCS pairing gap $\Delta$). The
process describes qualitatively the coupling of the hole excitation
to a gapless phonon mode.}
  \label{fig:altdiagrammi}
\end{figure}

The first term in Eq.~\eqref{eq:wavef2} can be depicted pictorially as a 
diagram in Fig.~\ref{fig:altdiagrammi} (the second term corresponds to
diagram in Fig.~\ref{fig:diagrammiother} as will be discussed later).
The diagram has
been drawn for a more general spectroscopy than rf-spectroscopy by
including also the photon momentum $q_\mathrm{L}$. Below
we will assume that $q_\mathrm{L} = 0$ for rf-spectroscopy, but
later on in the discussion regarding Bragg spectroscopy the momentum of
the photon can 
be significant. The diagram describes a path for the wavefunction and
one will need to calculate
the squared norm of the amplitude in order to find expectation values, 
such as a Green's function. As in the discussion 
regarding linear response above,
a first-order perturbation for the wavefunction corresponds to 
a second-order perturbation for the propagator.


\begin{figure}
  \centering
  \includegraphics[width=0.60\textwidth]{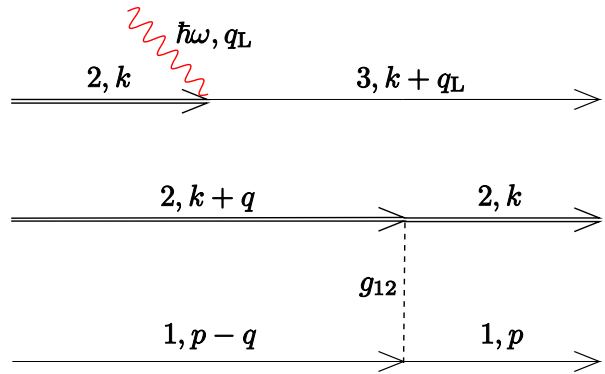}
  \caption{An alternative way of seeing the process in 
Fig.~\ref{fig:altdiagrammi}.  The absorption of the photon
by an atom in the ground state of $|2\rangle$ produces 
a particle in hyperfine state $|3\rangle$.
After the photon absorption, another pair of atoms in states $|2,k+q\rangle$
and $|1,p-q\rangle$ scatter into states $|2,k\rangle$, $|1,p\rangle$, 
filling the hole created by the photon absorption.} 
  \label{fig:diagrammi}
\end{figure}

In the case of rf-spectroscopy, the diagram describes a correlated process 
in which a rf-photon creates an atom in state $|3,k\rangle$ and a hole in 
$|2,k\rangle$. The hole then scatters with a hole in state $|1,p\rangle$ 
into a new set of holes $|2,k+q\rangle$, $|1,p-q\rangle$. Another way
to describe the diagram is that an atom in state $|2,k\rangle$ absorbs the
photon and is transferred into the state $|3,k\rangle$. 
After the rf-photon absorption, another pair of atoms, 
originally in states $|2,k+q\rangle$ and $|1,p-q\rangle$, scatter into 
states $|2,k\rangle$, $|1,p\rangle$, filling the hole created by the
RF-transition. This description can be better illustrated with diagram in 
Fig.~\ref{fig:diagrammi}. While the latter description sounds like the 
processes 
(rf-photon absorption and atom-atom scattering) are disconnected, 
the first interpretation shows that they are indeed connected 
through hole degrees of freedom. However, below we will use the latter
description as it generalizes more easily. The corresponding diagrams
for holes can be easily formulated as the two representations are completely
interchangeable.

In the following, we will assume that the propagator lines for atoms in 
hyperfine state $|1\rangle$ are described by bare Green's 
functions and only the atoms (or holes) in hyperfine state $|2\rangle$ 
are described by dressed propagators. If we look at the residual 
interactions $H_0^\mathrm{res}$ in the 
quasiparticle basis, it is clear that in the first order of scattering 
the single quasiparticle created by the 
RF-field couples only to a state with three quasiparticles 
(through Eq.~\eqref{eq:quasiham2}) 
and not to the gapless Anderson-Bogoliubov phonon with a single quasiparticle. 
This problem is generic to a perturbative
approach, as the description of the collective Anderson-Bogoliubov phonon
would require self-consistent, or nonperturbative, treatment.
The assumption invoked here allows us to capture elements from the 
Anderson-Bogoliubov phonon since the scattering
does \emph{not} introduce extra quasiparticles. Indeed, in the
zero-momentum scattering asymptote $q \rightarrow 0$, the energy shift
due to the scattering becomes $\hbar c q$, where $c = \frac{\hbar p}{m}$.
Close to the Fermi surface, $p \approx k_\mathrm{F}$, this yields the speed 
of sound as expected for the AB phonon.

Thus by using bare propagators for atoms in state $|1\rangle$, the 
diagram~\ref{fig:diagrammi} yields an approximative description
for the coupling to the Anderson-Bogoliubov mode. To elaborate, by allowing
the single quasiparticle excitation to scatter into lower energy states, the
hole wavefunction will spread over several momentum states leading into 
the decay of the excitation.
However, since the process is correlated with the rf-photon absorption
it is possible for the quasiparticle excitation to scatter also to a higher
energy state if the rf-photon has excess energy. Simply put, this means that
the excess energy of the rf-photon is transferred into extra kinetic energy
of atoms. However, since the rf-transition conserves momentum, this
change in kinetic energy is provided by the atom-atom scattering. 
Similarly, given extra energy from the probing photon, a quasiparticle
at the bottom of the quasiparticle energy band (corresponding to a hole 
at the Fermi surface) can 
decay by scattering to higher energy states, acquiring a finite lifetime.
This might be connected to the observation of a finite linewidth
of such quasiparticles in Ref.~\cite{Haussmann2009}.

The diagram~\ref{fig:diagrammi} can be added to the standard linear response 
calculation of the radio-frequency spectroscopy, thus generalizing the 
linear response calculation to 
include couplings with gapless phonon modes.
As discussed already in Ref.~\cite{Haussmann2009}, such coupling broadens 
the sharp quasiparticle peak in the hole spectral function. We will 
discuss the implications later on in the section~\ref{sec:experiments}.

Including the diagram in Fig.~\ref{fig:diagrammi} yields the rf-transition 
(for zero photon momentum $q_\mathrm{L} = 0$)
\begin{equation*}
\fl \int_0^\infty dt\, \int_0^t dt'\, e^{i (\delta E_1 + \delta E_2)t} \left( g_{12} c_{1,p}^\dagger c_{2,k}^\dagger c_{2,k+q} c_{1,p-q} \right) e^{-i \delta E_1 (t-t')} \left( \Omega c_{3,k}^\dagger  c_{2,k} \right) |\psi(0)\rangle e^{-\eta t'},
\end{equation*}
where $\delta E_1 = \epsilon_k^3 - \mu + E_k^2 - \delta$ is the energy 
change due to rf-photon absorption (describing the energy of a single 
particle in state $|k,3\rangle$ and hole in $|2,k\rangle$), 
$\delta E_2 = \epsilon_p^1 - \epsilon_{p-q}^1 + E_{k+q}^2 - E_k^2$ is the energy change due to the
scattering (corresponding to the kinetic energy difference of the atom in hyperfine state $|1\rangle$
and the change in the hole excitation energy due to the scattering of the hole from state $|2,k\rangle$
to state $|2,k+q\rangle$), and the factor $e^{-\eta t'}$ has been added to guarantee convergence. 
The factor $\eta$ plays the role of linewidth of the rf-field and it can be
interpreted as a switching off of the rf-field. Here we have included the 
hyperfine state 
indices in the single-particle and quasiparticle energies for clarity. 
The time integrals will yield the energy conservation but in order to 
calculate the actual transition probability amplitude, one will need to 
calculate also the norm of the state
\[
  |\psi\rangle = c_{1,p}^\dagger c_{2,k}^\dagger c_{2,k+q} c_{1,p-q} c_{3,k}^\dagger  c_{2,k} |\psi(0)\rangle,
\]
which depends on the details of the initial state $|\psi(0)\rangle$.
The transition probability can be obtained by calculating the norm of this 
transition. However, as will be seen later, 
there are also other diagrams describing the same transition and these 
will need to be summed coherently before calculating the probabilities.

The time integrals can be easily evaluated, yielding
\begin{eqnarray}
\fl   P_{\vec{k},\vec{p},\vec{q}}^{\mathrm{hole}} |\psi\rangle &= \frac{\Omega}{\delta E_1 + i\eta'} \frac{g_{12} c_{1,p}^\dagger c_{2,k}^\dagger c_{2,k+q} c_{1,p-q} c_{3,k}^\dagger c_{2,k}|\psi(0)\rangle}{\delta E_1 + \delta E_2 + i\eta} \nonumber \\
\fl &= \frac{\Omega}{\epsilon_k^3 - \mu + E_k^2 - \delta + i\eta'} \frac{g_{12} c_{1,p}^\dagger c_{2,k}^\dagger c_{2,k+q} c_{1,p-q} c_{3,k}^\dagger c_{2,k}|\psi(0)\rangle}{ \epsilon_p^1 + \epsilon_k^3 + E_{k+q}^2 - \mu - \epsilon_{p-q}^1 - \delta + i\eta},
\label{eq:scatspec3}
\end{eqnarray}
where we have added a lifetime broadening $\eta'$ to the (virtual) intermediate state.
We use the lifetime broadening 
$\eta' = |g_{12}|N_1$, where $N_1$ is the number of particles in state 
$|1\rangle$, because the two-body scatterings are the dominant decay channel in a 
dilute gas. 

For narrow linewidth $\eta \ll \eta'$, the transition probability amplitude
has a single peak centered around $\delta = (\epsilon_k^3 + \epsilon_p^1) - ( \mu-E_{k+q}^2 + \epsilon_{p-q}^1)$. This corresponds to the energy change
of a process in which one transfers one dressed ground state particle from 
state 
$|2,k+q\rangle$ and one bare particle from state $|1,p-q\rangle$ into two 
bare particles in states $|3,k\rangle$ and $|1,p\rangle$. Notice that
in the limit $q \rightarrow 0$ the peak approaches the standard linear
response result $\delta = \epsilon_k^3 - \mu + E_k^2$, the asymptotic
difference being linear in $q$ as already discussed above.

\section{Sum rule considerations}

\begin{figure}
  \centering
  \includegraphics[width=0.60\textwidth]{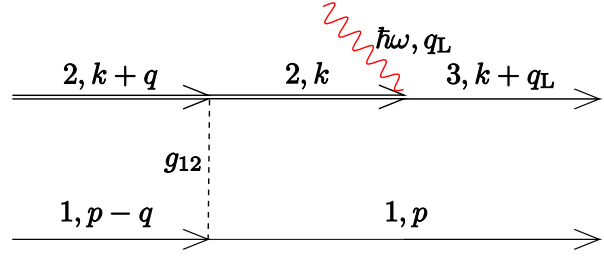}
  \caption{The other first-order diagram for the photon absorption 
$\hbar \omega,q_\mathrm{L}$ and atom-atom scattering $g_{12}$ process. 
Here the atom in hyperfine state $|2\rangle$ scatters, producing
one particle excitation and one hole excitation. The particle excitation
is then transferred to the hyperfine state $|3\rangle$ by the photon
absorption.}
  \label{fig:diagrammiother}
\end{figure}

\begin{figure}
  \centering
  \includegraphics[width=0.60\textwidth]{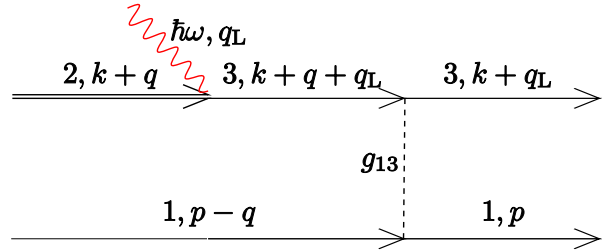}
  \caption{The first-order diagram for the rf-photon absorption 
$\hbar \omega,q_\mathrm{L}$ and for atom-atom scattering $g_{13}$ 
process. Here again the rf-photon creates a particle in hyperfine state
$|3\rangle$ and a hole excitation in $|2\rangle$. However, now the scattering
takes place in the $13$-channel. Assuming that initially the hyperfine
state $|3\rangle$ is empty, this is the only first-order diagram contributing
in this interaction channel.}
  \label{fig:diagrammi13}
\end{figure}

Based on very general arguments, the atom-atom interaction effects should 
vanish from the radio-frequency spectra when the interaction strengths between
atoms in hyperfine states $|1\rangle$ and $|2\rangle$ ($g_{12}$) and 
between states $|1\rangle$ and $|3\rangle$ ($g_{13}$) are 
equal~\cite{Yu2006a, BBaym2007a}. Thus we expect that the 
diagram in Fig.~\ref{fig:diagrammi} needs to be combined with other
diagrams of the same order in photon coupling and atom-atom interactions.
The other first-order diagrams are shown in Figs.~\ref{fig:diagrammiother} 
and~\ref{fig:diagrammi13}. The contributions from these diagrams are
\begin{equation}
\label{eq:scatspec}
\fl   P_{\vec{k},\vec{p},\vec{q}}^{12} |\psi\rangle= \frac{g_{12}}{E_k^2 + \epsilon_p^1 + E_{k+q}^2 - \epsilon_{p-q}^1 + i\eta'} 
\frac{\Omega c_{3,k}^\dagger c_{2,k} c_{1,p}^\dagger c_{2,k}^\dagger c_{2,k+q} c_{1,p-q} |\psi(0)\rangle}{\epsilon_p^1 + \epsilon_k^3 + E_{k+q}^2  - \mu - \epsilon_{p-q}^1 - \delta + i\eta},
\end{equation}
and
\begin{equation}
\label{eq:scatspec2}
\fl   P_{\vec{k},\vec{p},\vec{q}}^{13} |\psi\rangle= \frac{\Omega}{\epsilon_{k+q}^3 - \mu + E_{k+q}^2 - \delta + i\eta'} \frac{g_{13} c_{1,p}^\dagger c_{3,k}^\dagger c_{3,k+q} c_{1,p-q} c_{3,k+q}^\dagger c_{2,k+q} |\psi(0)\rangle}{\epsilon_{p}^1 + \epsilon_{k}^3 + E_{k+q}^2 - \mu - \epsilon_{p-q}^1 - \delta + i\eta}.
\end{equation}
We refer to the diagrams~\ref{fig:diagrammi},~\ref{fig:diagrammiother}, 
and~\ref{fig:diagrammi13} collectively by term resonant scattering and 
absorption (RSA) diagrams.

The quasiparticle line entering the scattering vertex in 
diagram~\ref{fig:diagrammiother} describes a 
(mean-field) ground state pair 
(energy $-E_{k+q}^2 < 0$) and the line leaving the scattering vertex is 
excited ($E_k^2 > 0$).
The role of the photon is now to transfer the particle excitation
to the state $|3\rangle$. Since quasiparticle has already been excited
in the atom-atom scattering process, the photon absorption itself 
does not create further excitations.
This follows also from the observation that the total energy difference
for all three diagrams is the same as they describe the same transition. 
The two diagrams shown in Figs.~\ref{fig:diagrammi} 
and~\ref{fig:diagrammi13} do not create more than one quasiparticle
excitation and hence the process in Fig.~\ref{fig:diagrammiother} cannot 
result in more than one excitation.

The diagram~\ref{fig:diagrammiother} warrants even a closer examination.
It describes a process in which the atoms scatter 
\emph{before} the photon is absorbed. The first part of the diagram
(the scattering) can be understood as a first-order correction to the 
mean-field ground state. Indeed, neglecting the photon line for a second,
the state obtained from the first-order time-dependent perturbation theory,
when the perturbation (in this case the atom-atom interaction) is switched
on adiabatically, is the state one would have obtained from the
first-order time-\emph{independent} perturbation theory. The perturbative
correction to the state can be understood as a fluctuation around the 
unperturbed state, and the photon in the diagram~\ref{fig:diagrammiother} 
couples to these fluctuations. 
While the diagram~\ref{fig:diagrammiother} could be removed by including
the first-order time-independent perturbative correction to the initial state
$|\psi(0)\rangle$ (that is, by including the fluctuations in the state $|\psi(0)\rangle$), the diagram in Fig.~\ref{fig:diagrammi} cannot. If we interpret the 
diagram~\ref{fig:diagrammiother} as a process in which fluctuations absorb
the photon, the diagram~\ref{fig:diagrammi} corresponds to a process
in which the fluctuations are \emph{enhanced} by the absorption of the photon.
Clearly, the latter process cannot be included in the initial state 
$|\psi(0)\rangle$ as the photon field is absent at time $t=0$. The present
formulation as time-dependent perturbation theory allows an easy and
consistent (in the sense of the sum rules discussed below) way to include
the effect of fluctuations in the spectroscopic probes.

\subsection{Asymptotic high energy scaling}

The form of Eq.~\eqref{eq:scatspec} has interesting implications for
the asymptotic tail of the radio-frequency spectrum. In a uniform
three-dimensional system the single-particle energy dispersion is
$\epsilon_k = \frac{\hbar^2 k^2}{2m}$. In the high energy asymptote of
the spectrum $\delta \gg E_\mathrm{F}$, pairing effects described by the
gap $\Delta$ have little 
effect (allowing the approximation 
$E_k^2 = \sqrt{\left(\epsilon_\vec{k} - \mu \right)^2 + \Delta^2} \approx \epsilon_\vec{k}$) 
and resonant processes are dominated by scatterings to high momentum 
states $p,k \gg k_\mathrm{F}$, where $k_\mathrm{F}$ is the Fermi momentum. 
Assuming that only low momentum states are initially occupied one
has $k+q \approx p-q \approx 0$ (hence $k \approx -p$ and $q$ is large). 
Now Eq.~\eqref{eq:scatspec} yields
\begin{equation}
   P_{\vec{q},\vec{k},\vec{p}}^\mathrm{12-asymptote} |\psi\rangle= \frac{g_{12}}{2\frac{\hbar^2k^2}{2m} + i\eta'} 
\frac{\Omega c_{3,k}^\dagger c_{2,k} c_{1,p}^\dagger c_{2,k}^\dagger c_{2,k+q} c_{1,p-q} |\psi(0)\rangle}{2\frac{\hbar^2k^2}{2m} - \delta + i\eta}.
\end{equation}
In the large $k$ limit, the probabilities for the states 
$|2,k\rangle$ and $|1,p=-k\rangle$ to be empty approach unity and 
likewise the occupation probabilities for states $|1,p-q\approx 0\rangle$ and 
$|2,k+q \approx 0\rangle$ approach unity.
Calculating the transition probability $|P_{\vec{q},\vec{k},\vec{p}}^\mathrm{12-asymptote} |\psi\rangle |^2$ and
integrating over $\vec{k}$ one obtains the transition 
probability for a single particle in state $|2,k+q\rangle$ to be 
transferred to hyperfine state $|3,k\rangle$
\begin{equation}
  P^\mathrm{12-asymptote} \sim \int_0^\infty d\epsilon \, \sqrt{\epsilon} \left|\frac{g_{12}}{2\epsilon + i\eta'} \frac{\Omega}{2\epsilon-\delta + i\eta}\right|^2,
\end{equation}
where the $\sqrt{\epsilon}$ factor comes from the density of states in a 
three-dimensional uniform system.
For narrow linewidth $\eta \ll \delta$, only energies $\epsilon$ close to $\delta/2$
contribute to the integral and it can be approximated as
\begin{equation}
\fl  P^\mathrm{12-asymptote} \sim \int_0^\infty d\epsilon \, \sqrt{\delta/2} \left|\frac{g_{12}}{\delta + i\eta'} \frac{\Omega}{2\epsilon-\delta + i\eta}\right|^2 = \sqrt{\delta/2} \frac{\left(g_{12}\Omega \right)^2}{(\delta^2 + \eta'^2)} \frac{\pi}{\eta}.
\label{eq:12asymptote}
\end{equation}
Assuming $\delta \gg \eta'$, we obtain the asymptotic high RF detuning tail 
$P_\vec{k}^\mathrm{asymptote} \sim \delta^{-1.5}$.
It is important to notice that this result is independent of a possible 
'quasiparticle tail' in the occupation numbers $v_{2,\vec{k}}$.
Diagrams~\ref{fig:diagrammi} and~\ref{fig:diagrammi13} also produce
high energy tails for the spectra. The asymptotic form of the
latter diagram is
\begin{equation}  
P^\mathrm{13-asymptote} \sim \int_0^\infty d\epsilon \, \sqrt{\epsilon} \left|\frac{\Omega}{-\delta + i\eta'} \frac{g_{13}}{2\epsilon-\delta + i\eta}\right|^2,
\end{equation}
which yields the same asymptotic behaviour as Eq.~\eqref{eq:12asymptote}
but is proportional to $g_{13}$ instead of $g_{12}$. 
Unlike the above two asymptotic forms, the high energy asymptote of 
diagram~\ref{fig:diagrammi} requires occupation of high momentum states and 
it is given by
\begin{equation}  
P^\mathrm{hole-asymptote} \sim \int_0^\infty d\epsilon \, \sqrt{\epsilon} \left|\frac{\Omega}{2\epsilon-\delta + i\eta'} \frac{g_{12} v_{2,\epsilon}}{2\epsilon-\delta + i\eta}\right|^2,
\label{eq:holeasymptote}
\end{equation}
where $v_{2,\epsilon}$ is the probability amplitude that there is an
atom in hyperfine state $2$ with energy $\epsilon$. For a BCS-like state
this is given by the Bogoliubov coefficient 
$v_{2,\epsilon}^2 \approx \frac{1}{2} \left( 1 - \epsilon/\sqrt{\epsilon^2 + \Delta^2} \right)$ yielding the same asymptotic high energy tail $\delta^{-1.5}$.
Here we have neglected terms of the order $v_k^4$ which is justified
assuming that the occupation probabilities do not have too fat a tail.
The most important point to notice, however, is that only
Eq.~\eqref{eq:holeasymptote} depends on high energy excitations in the initial
state $|\psi(0)\rangle$.

The $\delta^{-1.5}$ scaling law has been suggested already in various 
studies~\cite{Tan2008a,Tan2008b,Braaten2008a,Schneider2010a}, where the
scaling law comes from the high energy tail of the occupied atom states
(the $v_{2,k}$ or $v_{2,\epsilon}$ tail).
In present theory, no such many-body correlations are needed 
in the initial state -- the power law arises directly from the 
RSA process. 
In particular, the RSA scaling law applies in the limit 
$\Delta \rightarrow 0$ and also to all momentum $k$ modes separately,
resulting in a $\delta^{-1.5}$ tail in the spectral function of every
$k$ mode. This will have implications in particular for the momentum
resolved spectroscopy~\cite{Stewart2008a} that we will consider 
in Section~\ref{sec:momres}.

All the RSA diagrams lead into the same final state 
$|\psi\rangle$ and thus they interfere
coherently. A closer
examination of the orderings of the creation and annihilation operators 
shows that, because of fermionic anticommutation rules,  Eq.~\eqref{eq:scatspec2} has opposite sign to the 
Eqs.~\eqref{eq:scatspec3} and~\eqref{eq:scatspec}, and hence the three
contributions interfere destructively. This is easy to see by operating
on Eq.~\eqref{eq:scatspec2} with the identity operator $1 = c_{2,k}^\dagger c_{2,k}^{} + c_{2,k}^{} c_{2,k}^\dagger$.
In particular, for identical interactions ($g_{12} = g_{13}$) and for 
equal dispersions ($E_k^2 = \epsilon_k^3$, for all $k$), the three 
RSA diagrams combine
into a single \emph{sequential} process. Such process does not allow energy
exchange between the quasiparticle and the emitted phonon. Thus,
in the symmetric case, our model does not provide any energy shift to the
rf-spectrum. This is in agreement with sum rules.

The sum rule argument must hold for all orders of atom-atom scatterings, 
and hence it will hold also (or especially) in the nonperturbative treatment.
Thus the picture from our model, in which the rf-photon absorption and
the excitation of the phonon are correlated, should also hold for a real
system in which the phonon is the Anderson-Bogoliubov phonon.

If the final state $|3\rangle$ is interacting, also the BCS mean-field 
response changes and one needs to include the Aslamazov-Larkin (AL) and
Maki-Thompson (MT) contributions. These resemble our beyond mean-field 
diagrams. However, these are strictly mean-field effects and
do not contribute to the quasiparticle excitation lifetimes. The
effect of these diagrams have been studied in the context of
RF spectroscopy of atomic gases in Refs.~\cite{He2009a,Pieri2009a}.

\section{Comparison to experimental results: fermionic gases}
\label{sec:experiments}

All spectroscopies used in the context of ultracold atom gases can be 
understood as an effective photon absorption and hence we expect diagrams 
similar to those in Fig.~\ref{fig:diagrammi} to play a role. 
The main 
difference in various spectroscopic methods is the atom degree of freedom
that is affected: hyperfine state (RF and Raman spectroscopies), momentum 
state (Bragg and Raman spectroscopies), or lattice band 
(lattice modulation and Bragg spectroscopies). Furthermore, the diagrams
(with slight modifications) can be applied also to bosonic atoms.
In the following we will analyze seven different 
experiments which we believe to show the effect of the RSA process.

\subsection{Radio-frequency spectroscopy of a Fermi gas}
\label{sec:momres}

An experiment performed in JILA in 2008~\cite{Stewart2008a,Chen2009a} 
applied momentum resolved RF spectroscopy to probe the single-particle
excitation spectrum of a two-component Fermi gas. 
The spectrum in Fig.3c of Ref.~\cite{Stewart2008a} was shown to be
broader than expected and the effect was assumed to follow from the 
center-of-mass motion of the pairs in the BEC side of the Feshbach resonance.
However, in a 3D Fermi gas, also the RSA process produces the broadening 
effect. 
Furthermore, Fig.4 in Ref.~\cite{Stewart2008a} shows
asymmetric spectra for atoms of a given momentum. For a fixed momentum
$k$ the linear response theory predicts a symmetric spectrum centered
at the single-particle resonance at $\xi_k - \mu + E_k + U_{12} n_1$. 
The RSA diagram allows the transition with more energetic RF photons, 
producing a tail on the negative side of the single-particle peak. 

\begin{figure}
\centering
\includegraphics[width=0.90\textwidth]{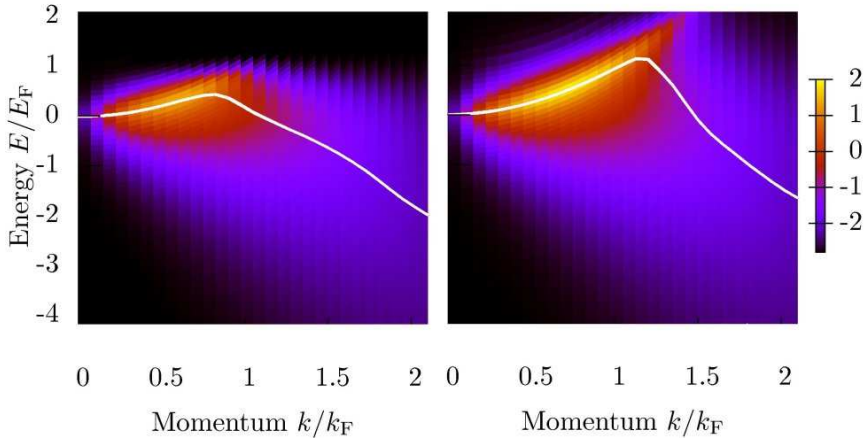}
\caption{The momentum resolved rf-spectra for the gapless system at 
zero temperature (left) and for $T = 0.2\,T_\mathrm{F}$ (right).
For low momenta $k \ll k_\mathrm{F}$, the spectrum follows the free
particle dispersion. For high momenta $k \gg k_\mathrm{F}$ the spectrum
follows a negative $-\frac{\hbar^2 k^2}{2m}$ dispersion. The transition 
region around $k_\mathrm{F}$ shifts towards higher momenta for increasing
temperature in qualitative agreement with the experimental results.
The white curves indicate the centers of gaussian fits to the individual
spectra for different momenta $k$. These are to be compared with the
experimental data in Refs.~\cite{Stewart2008a,Gaebler2010a}.}
\label{fig:momresspectra}
\end{figure}

The present theory can be easily adapted for describing this experiment.
Due to high temperatures, we assume that the initial state can be
described by a normal finite temperature Fermi gas, implying that the
quasiparticle energies in Eqs.~\eqref{eq:scatspec3} and~\eqref{eq:scatspec}
can be replaced by single particle dispersion $E_k^2 = |\epsilon_k-\mu|$.
The contributions from the two equations can be combined and we obtain
in the limit of narrow linewidth ($\eta \rightarrow 0$) the transition 
probability
\begin{equation}
\fl   |P_{k,p,q} |\psi\rangle|^2 = \frac{\Omega^2 g_{12}^2 n_\mathrm{F} (\epsilon_{p-q}) n_\mathrm{F}(\epsilon_{k+q}) n_\mathrm{F} (-\epsilon_p)}{(\delta)^2 + \eta'^2} \delta( \epsilon_{k+q} + \epsilon_p - \epsilon_k - \epsilon_{p-q}  - \delta),
\label{eq:momrescurrent}
\end{equation}
where $\delta (x)$ is the Dirac delta function and $n_\mathrm{F}(x) = 1/(1+e^{\beta(x-\mu)})$ is the Fermi-Dirac distribution at temperature $T = 1/k_\mathrm{B}\beta$, where $k_\mathrm{B}$ is the Boltzmann constant. The lifetime broadening $\eta'$ should in principle depend on the energy of the excitation, but here 
we will use the average lifetime obtained from the Galitskii 
equation~\cite{FetterWalecka}
\begin{equation}
  \frac{\eta}{E_\mathrm{F}}  = \frac{2\left(k_\mathrm{F}a\right)^2}{\pi} \frac{\int d^3 p\, \left( 1-\frac{|p|}{k_\mathrm{F}}\right)^2 \theta (k_\mathrm{F}-|p|)}{\int d^3 p \theta (k_\mathrm{F}-|p|)} = \frac{4}{5\pi},
\end{equation}
where we have replaced $k_\mathrm{F} a = -2$. This value is motivated by 
two-body scattering calculation which yields $k$-dependent scattering 
amplitude~\cite{FetterWalecka}. It should be noticed that the
qualitative results do not depend sensitively on the choice of this lifetime 
broadening. The momentum resolved spectra can be calculated
by summing over momenta $p$ and $q$ in Eq.~\eqref{eq:momrescurrent} and 
plotting the transition probabilities as functions of momentum $k$ and 
energy $\epsilon_k-\delta$. Doing this we obtain the spectra shown in 
Fig.~\ref{fig:momresspectra}, which can be directly compared with momentum 
resolved spectra in 
Refs.~\cite{Stewart2008a,Gaebler2010a}. Notice, however, that the latter 
experiment is performed at temperatures below the critical temperature, 
and it is somewhat surprising that in this case we obtain good match
with the experimental spectra, having assumed a gapless state.
This warrants
a more detailed analysis, which will be a topic of further work. Despite the 
simplicity of the current model, the qualitative agreement is excellent. 
Clearly, normal state correlations
are sufficient for producing the back-bending $-\epsilon_k$ dispersion,
as also seen in Ref.~\cite{Schneider2010a} where a more complete
T-matrix calculation was carried out for $k \gg k_\mathrm{F}$. The present
model, however, can describe also the low momentum regime where the
hole degrees of freedom become important. 

It is worth pointing out that possibly the most controversial assumption
made in our theory, namely assuming that the atoms in hyperfine
state $|1\rangle$ are described by bare propagators, was not needed in the
above calculation of the momentum resolved spectrum for a normal state
since the dressed and bare propagators are identical.

In a recent experiment by the same group~\cite{Stewart2010a} the high 
momentum asymptote
was studied also by measuring the expansion of the atom gas after abruptly 
switching off the interactions. This asymptote was found to match very
well with the high momentum tail obtained from the rf-spectrum, in agreement
with the theory of S. Tan~\cite{Tan2008a, Tan2008b}. It is worth noticing
the similarity between transferring a particle into a noninteracting
hyperfine state $|3\rangle$ and switching off of the interactions. While
in the first the moment of the transition is not well defined (reflected
as a time integral in the response), in the latter the moment when
the interactions are switched off is determined by the external magnetic field.
Despite this important difference, one can apply the RSA model also to the
latter setting. The only contributing diagram is the 
Fig.~\ref{fig:diagrammiother} as the scattering must take place before the
interactions are switched off. Now the corresponding integral formula is
\begin{equation}
\fl \frac{i}{\hbar} \int_0^T dt\, e^{-i \left( E_k^2 + \epsilon_p^1 + E_{k+q}^2 - \epsilon_{p-q}^1 \right) (T-t)/\hbar} \left( g_{12} c_{1,p}^\dagger c_{2,k}^\dagger c_{2,k+q} c_{1,p-q} \right) |\psi(0)\rangle e^{-\eta t}.
\end{equation}
Evaluating this equation, and taking the limit $T \rightarrow \infty$ (one 
could also keep the time $T$ finite but we want to compare the result
with rf-spectra for which we considered the limit of infinitely long pulse),
yields
\begin{equation}
 \frac{g_{12}}{E_k^2 + \epsilon_p^1 + E_{k+q}^2 - \epsilon_{p-q}^1 + i \eta} c_{1,p}^\dagger c_{2,k}^\dagger c_{2,k+q} c_{1,p-q} |\psi(0)\rangle.
\end{equation}
Now the probability of finding an atom in hyperfine state $|2\rangle$ with
momentum $k$ is
\begin{equation}
 \left| \sum_{p,q} \frac{g_{12}}{E_k^2 + \epsilon_p^1 + E_{k+q}^2 - \epsilon_{p-q}^1 + i\eta} c_{1,p}^\dagger c_{2,k}^\dagger c_{2,k+q} c_{1,p-q} |\psi(0)\rangle \right|^2.
\end{equation}
As above, in the discussion of the high energy tail of the rf-spectrum,
we obtain constraints $p-q \approx k+q \approx 0$ and the energy change
due to the scattering becomes
$E_k^2 + \epsilon_p^1 + E_{k+q}^2 - \epsilon_{p-q}^1 \approx 2 \epsilon_k$,
yielding the \emph{perceived occupation probability} for momentum state $k$
in the high momentum $k \rightarrow \infty$ limit 
\begin{equation}
 \left| \frac{g_{12}}{2\epsilon_k + i\eta} \right|^2 \sim \frac{1}{k^4}.
\end{equation}
This $k^{-4}$ occupation probability produces the $\delta^{-3/2}$ asymptote
in the rf-spectrum in linear response theory and this asymptotic behaviour
is also exactly the same as obtained for rf-spectrum in RSA model.
Thus the tail observed in Ref.~\cite{Stewart2010a} is also in agreement with the
RSA model. However, we have made no assumptions regarding the occupation 
numbers for the initial state $|\psi(0)\rangle$.

\subsection{Bragg spectroscopy of a Fermi gas}

Another experiment used Bragg spectroscopy for studying a strongly interacting
Fermi gas~\cite{Veeravalli2008a}. The experiment fitted very well with 
the theoretical mean-field predictions including both the heights and 
the widths of the spectral peaks. 
In the case of Bragg spectroscopy,
the hyperfine state of the atom is not changed in the photon absorption
but only the momentum state. Therefore, the hyperfine state $|3\rangle$
in the RSA diagrams corresponds to hyperfine state $|2\rangle$. 
Thus the final state interactions
discussed in the context of rf-spectroscopy match
the initial interactions $g_{13} = g_{12} = g$, and the diagrams 
interfere destructively. Thus one does not expect any contribution
from the RSA effect to the Bragg spectroscopy in a Fermi gas. This is in 
good agreement with the experiment.
This observation shows that the Bragg spectroscopy could be a good way
for measuring single-particle properties in a Fermi gas since the
RSA-type beyond mean-field effects are suppressed.

\section{Comparison to experimental results: bosonic gases}

The RSA effect can be applied also to bosonic atoms. However, especially
in the presence of a Bose-Einstein condensate, the effect of different
diagrams is changed because atoms within the condensate do not feel the
exchange interaction channel~\cite{Pethick2001a}. In the following, we will
first consider a superfluid Bose gas and experiments performed on such 
systems, and later on discuss the Mott insulator state and the corresponding
experiments.

\subsection{RSA effect in BEC}

Let us consider a single-component Bose gas and let us assume, for the 
sake of simplicity, that all atoms are initially Bose-Einstein condensed 
into a zero-momentum state.
Now the relevant RSA diagrams for Bragg and lattice modulation spectroscopies 
are the ones shown in Figs.~\ref{fig:diagrammiother} and~\ref{fig:diagrammi13},
but since there is only one hyperfine component present, the hyperfine states
$|1\rangle$, $|2\rangle$, and $|3\rangle$ are the same.
The diagram in Fig.~\ref{fig:diagrammi} does not contribute in the case 
of an ideal BEC because the diagram would require atoms initially in two 
different momentum states 
$k$ and $k+q$ and hence requires uncondensed atoms. Of course, such 
process would be allowed for non-ideal BEC and may be important in the case
of strongly interacting BECs where quantum depletion is important or at high
temperatures.

As in the case of fermions, these bosonic RSA diagrams interfere coherently,
but the only contribution comes from momenta $k+q = p-q = 0$. However,
the diagram~\ref{fig:diagrammi13} experiences also the exchange
interaction channel as the atom-atom scattering does not take place between
two condensed atoms but between one condensed atom and one atom with momentum 
$q_\mathrm{L}$. Hence the two RSA diagrams do not fully cancel each
other. This partial cancellation of the RSA contribution produces 
interesting effects in the case of a superfluid BEC in an optical lattice.

\subsection{Lattice modulation spectroscopy of a Bose gas}

\begin{figure}
  \centering
    \includegraphics[angle=-90,width=0.60\textwidth]{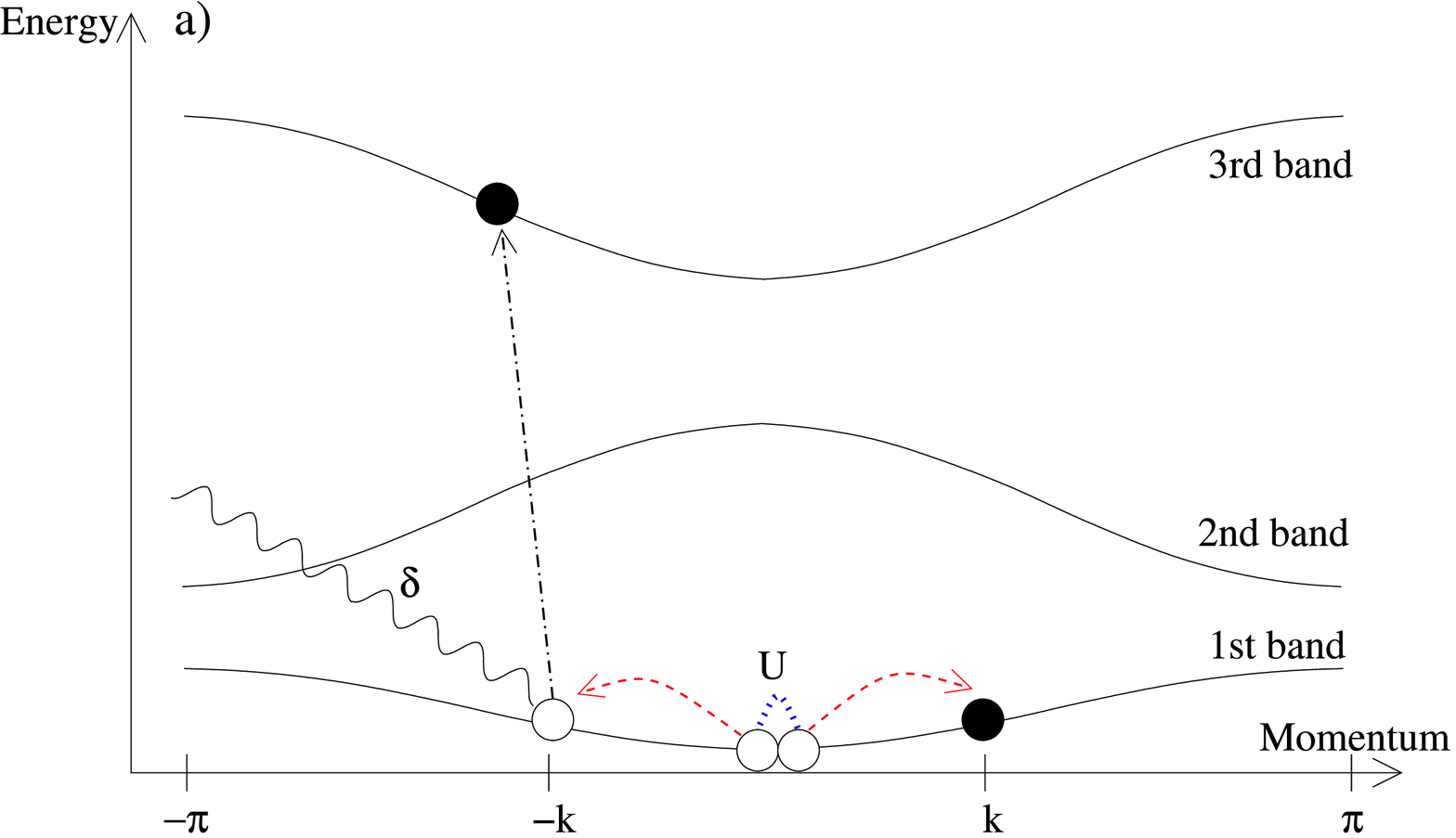}\\
    \includegraphics[angle=-90,width=0.60\textwidth]{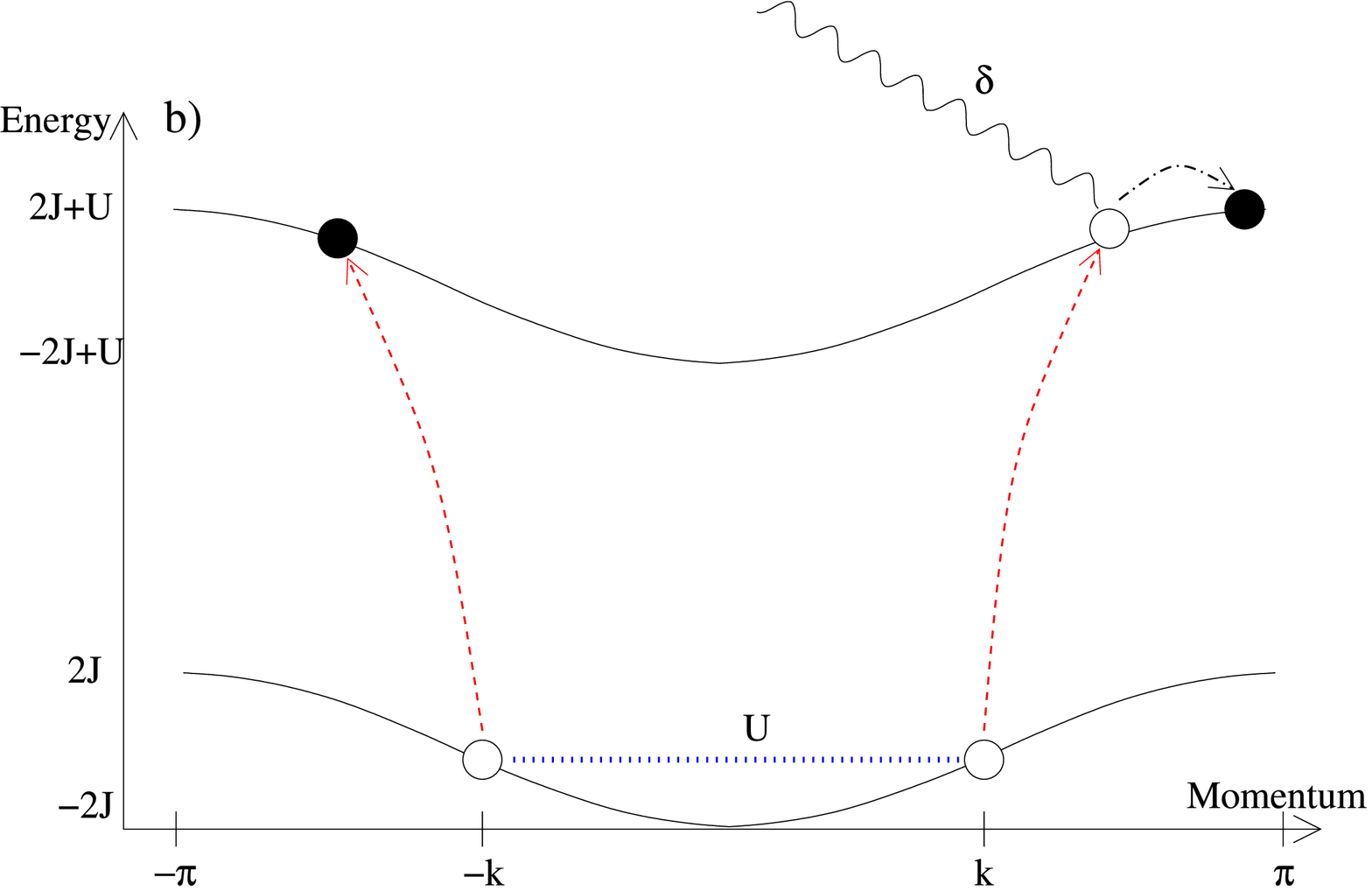}
  \caption{Bosonic RSA energy level scheme relevant for Bragg scattering and 
lattice modulation experiments~\cite{Fabbri2009a},~\cite{Clement2009a}, 
and~\cite{Stoferle2004a}. a) The atom-atom interaction $U$ excites two 
particles (two dashed arrows) from the zero-momentum condensate. One of 
the atoms is then scattered by the photon $\delta$ into the third band 
(the single dot-dashed line). b) In the Mott state, 
the atom-atom interaction $U$ will yield a large energy shift due to the 
Mott gap, leading into a two-peak structure in the spectrum. The Bragg photon
$\delta$ provides an additional momentum kick (dot-dashed line) and needs to 
provide the energy for creation of the two Mott excitations.}
  \label{fig:bragg}
\end{figure}

A superfluid Bose gas in a 1d lattice was studied using lattice 
modulation spectroscopy~\cite{Stoferle2004a} and Bragg
spectroscopy~\cite{Fabbri2009a,Clement2009a}. In the
experiments the atoms were transferred within the first lattice
band~\cite{Stoferle2004a,Clement2009a} as well as between the first lattice 
band and the second and third bands~\cite{Fabbri2009a}, while changing the
atom momentum due to the momentum carried by the photons.
The transition to the second band was as predicted by the standard Bogoliubov
(mean-field) theory, but the spectral peak of the transition to the third band 
and transitions within the first band were anomalously broad (see Fig.3
in~\cite{Fabbri2009a}, Fig.3a in~\cite{Clement2009a}, and 
Fig.1 in~\cite{Stoferle2004a}).
These effects were attributed to a loss of coherence~\cite{Fabbri2009a}, 
quantum depletion of the condensate~\cite{Stoferle2004a}, and to beyond 
mean-field correlation and thermal effects in a 1d system~\cite{Clement2009a}.

The present theory suggests an alternative explanation. Since atoms in
different lattice bands have reduced overlap, the interband interactions
are reduced. In the harmonic potential approximation, atoms in the second 
band have an extra factor $0.5$ in the interactions with the first band
(and in particular with the condensate), and the atoms in the third band have a
factor $0.375$. Including these factors, the bosonic RSA diagrams
will cancel each other for the transition to the second band (as the
contribution from the diagram~\ref{fig:diagrammi13} is reduced
by factor $0.5$, cancelling the exchange interaction channel effect) but 
will yield only a factor $1-2 \cdot 0.375=0.25$ 
(contribution
from diagram~\ref{fig:diagrammi13} is multiplied by $0.375$)
for the transition to the third band and a factor $1$ for the 
transitions within the first band.
Thus the bosonic RSA effect does not affect the transitions to the second
lattice band but it is present for the transitions to the first and the third
bands. Similarly to the fermionic gases, the bosonic RSA effect will spread
the spectral peaks. This is in qualitative agreement with the experimental
observations. Furthermore, the width of the transition to the third band 
approaches the width of the third band and the width of the transitions 
within the first band is approximately twice the first band width, in 
agreement with present theory. Fig.~\ref{fig:bragg} a) shows the schematic 
energy level diagram for the transition to the third band.

\subsection{Bragg spectroscopy of a Bose gas}

The Bragg spectroscopy of a Bose gas has been studied also in a harmonic
trap. The experiment in JILA in 2008 applied Bragg spectroscopy to study a 
strongly interacting BEC~\cite{Papp2008a,Kinnunen2009a,Ronen2009a}. 
The width of the spectral peak showed anomalous broadening 
in the strongly interacting regime (see Fig.3 in~\cite{Papp2008a}).
The bosonic RSA process explains the broadening by allowing
transitions at higher energies as in the above discussion regarding the
1d BEC experiments~\cite{Stoferle2004a,Fabbri2009a,Clement2009a}
and the momentum-resolved rf-spectroscopy~\cite{Stewart2008a}. 
Moreover, the bosonic RSA process provides transitions also at lower energies, 
since non-condensed (at finite momentum $k$) atoms can be scattered into 
the condensate before absorbing the Bragg photon (in which case the 
atom-atom scattering before the photon absorption feels the exchange 
interaction channel but the opposite process does not). This reduces 
the kinetic energy of the atoms, 
allowing transitions at lower frequencies and thus further contributing to the 
width of the spectrum. These atom-atom scatterings down into the condensate
are enhanced by the coherent state, and thus both RSA processes, scatterings 
up from
the condensate and scatterings down into the condensate, are equally likely
even when
the quantum depletion is weak. Since the first process produces a high 
energy tail due to 
transitions to higher kinetic energy states, the latter process produces 
a similar low
energy tail. Hence the total spectral lineshape remains symmetric, as 
also observed
in the experiment~\cite{Papp2008a}.

Notice that a similar process as described by the RSA effect was crucial 
for the experiment, as the actual measured response was the total 
momentum of the atom gas~\cite{Papp2008a}. The single particle excitations 
created by the Bragg scattering
decayed due to scatterings with other atoms and the excess momentum
yielded by the Bragg photon was thus transferred into center-of-mass momentum
of the atom cloud. Thus the single-particle excitation was turned into
a collective motion of the cloud. What our RSA effect now suggests, is
that these atom-atom scatterings and the initial Bragg scattering are 
coherent in the sense that they are able to exchange energy as well as 
momentum.

\subsection{RSA effect in bosonic Mott insulator}

In the case of a Mott insulator state in a Bose gas in an optical lattice
also the diagram~\ref{fig:diagrammi} contributes as the condensate fraction 
vanishes and higher momentum states become populated in the ground state.
Since there is no condensate present, all interactions experience the 
exchange interaction channel. However, the biggest difference to the 
interpretation of these diagrams as compared to superfluid gases is that 
the excitations created by the photon absorption do not couple
to gapless excitations, as the gapless excitations in the Mott state 
are spin-flip excitations and these are not induced by atom-atom scatterings. 
Instead, the scattering of two atoms in the Mott 
insulator will necessarily create two single-particle excitations
that are gapped by the Mott gap $U$, where $U$ is the on-site interaction
energy. The energy level scheme of the bosonic 
RSA process in a Mott insulator is shown in 
Fig.~\ref{fig:bragg} b), showing how the process produces two Mott excitations.

The 1d Bose gases in the Mott insulator regime were studied
using Bragg spectroscopy~\cite{Clement2009a} and lattice modulation 
spectroscopy~\cite{Stoferle2004a}. 
The measured spectra, such as in Figs.1 f)-h) in~\cite{Clement2009a} and
Fig.1 in~\cite{Stoferle2004a}, show two distinct peaks, the second
resonance at twice the energy of the first resonance.
The second peak was attributed to defects in the lattice, with atoms
being transferred to doubly occupied sites~\cite{Stoferle2004a}.
These observations fit in the picture drawn above as the additional peaks
observed in~\cite{Clement2009a} and~\cite{Stoferle2004a} can be
understood as a creation of two Mott excitations with energy cost
equal to twice the Mott gap $g$. The process is made resonant
by sufficiently energetic photon absorption, reflected as the second
peak in the spectra. Notice that in the second order of atom-atom
scatterings, one can also create three Mott excitations. Such processes
should result in spectral peaks at three times the Mott gap.

\subsection{Afterword}

To summarize, we analysed seven experiments in the light of our theory,
and found out that there is a qualitative match. However, for the case of
momentum resolved rf-spectroscopy we performed quantitative analysis, obtaining
good agreement with the experiment and the theory.
As in the case of momentum resolved spectroscopy in Fermi gases, calculating 
quantitative predictions for also the other experiments using RSA-theory can
be done. For example in the case of 
spectroscopies of superfluid BEC's in 
Refs.~\cite{Papp2008a,Stoferle2004a,Fabbri2009a,Clement2009a}, the starting 
point could be an ideal BEC and the interactions would then
be considered only in the first-order perturbation theory by the present
diagrammatic method. The approximation can be also easily improved by
using the Bogoliubov theory, and the subsequent ground state, as the starting
point. This is, indeed, an interesting topic for further work.

\section{Single-band Fermi Hubbard model}
\label{sec:tebd}

Besides experiments, also exact numerical results provide a good
platform for testing the theory. In this section we will study  
a Fermi gas in a one-dimensional optical lattice described 
by the single-band Fermi-Hubbard model. We will compare the predictions 
from the RSA theory to the exact rf-spectra obtained
using time-evolved block decimation (TEBD) method.

In the language used in these systems, the radio-frequency photon 
can be understood as creating
a spin excitation and the RSA process describes then the decay
of this excitation by scattering into lower energy spin excitation and
creating a gapless charge excitation. The charge excitation thus plays
the role of the Anderson-Bogoliubov phonon.
We apply the time-evolved block decimation (TEBD) method as it allows us to 
access the exact radio-frequency spectrum of the system.
The time-evolution of the system is determined by the Fermi-Hubbard
Hamiltonian
\begin{equation}
\fl  {H}(t) = -J\sum_{i,\sigma \in \{1,2,3\}} {c}_{i \sigma}^\dagger {c}_{i+1 \sigma} + H.c. + 
	U \sum_i {n}_{i2} \, {n}_{i1}
 + \Omega e^{i\delta t/\hbar} \sum_i {c}_{i 3}^\dagger {c}_{i 2} + H.c
	\label{eq:hubbard}
\end{equation}
where $U<0$ is the attractive on-site interaction strength between atoms in 
states $|1\rangle$ and $|2\rangle$  and the density operator 
${n}_{i\sigma} = {c}_{i\sigma}^\dagger {c}_{i\sigma}$, 
where ${c}_{i\sigma}^{(\dagger)}$ 
destroys (creates) a particle of hyperfine state $|\sigma\rangle$ in 
lattice site $i$. The state $|3\rangle$ is assumed to be noninteracting.
Below we use units that have the hopping strength $J = 1$.

\begin{figure}
\centering
  \includegraphics[width=0.60\textwidth]{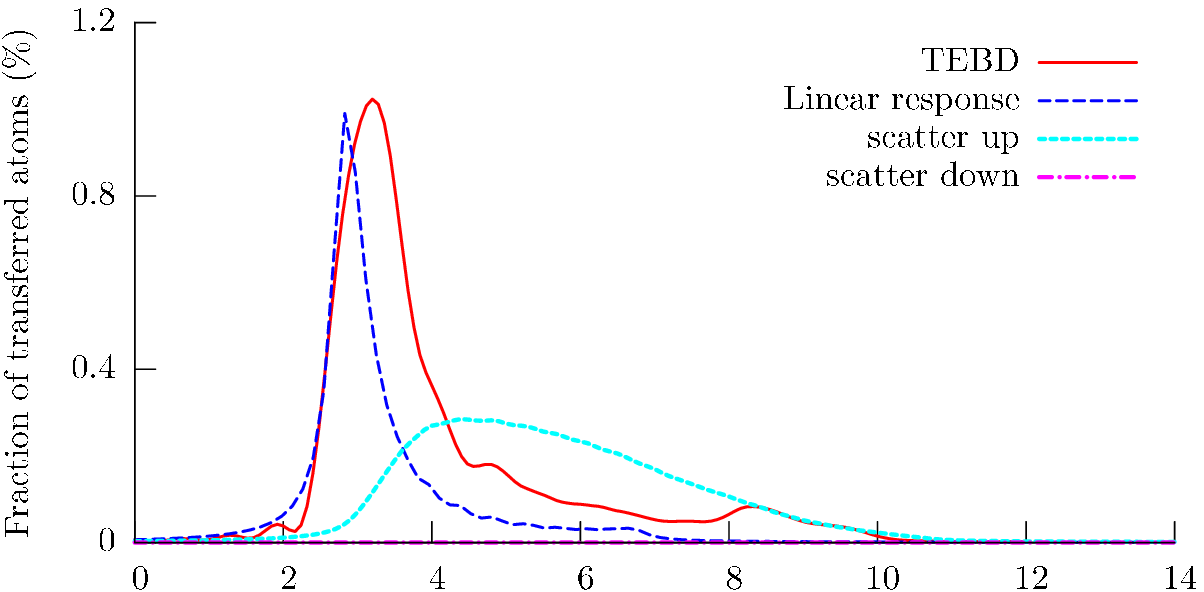}\\
  \includegraphics[width=0.60\textwidth]{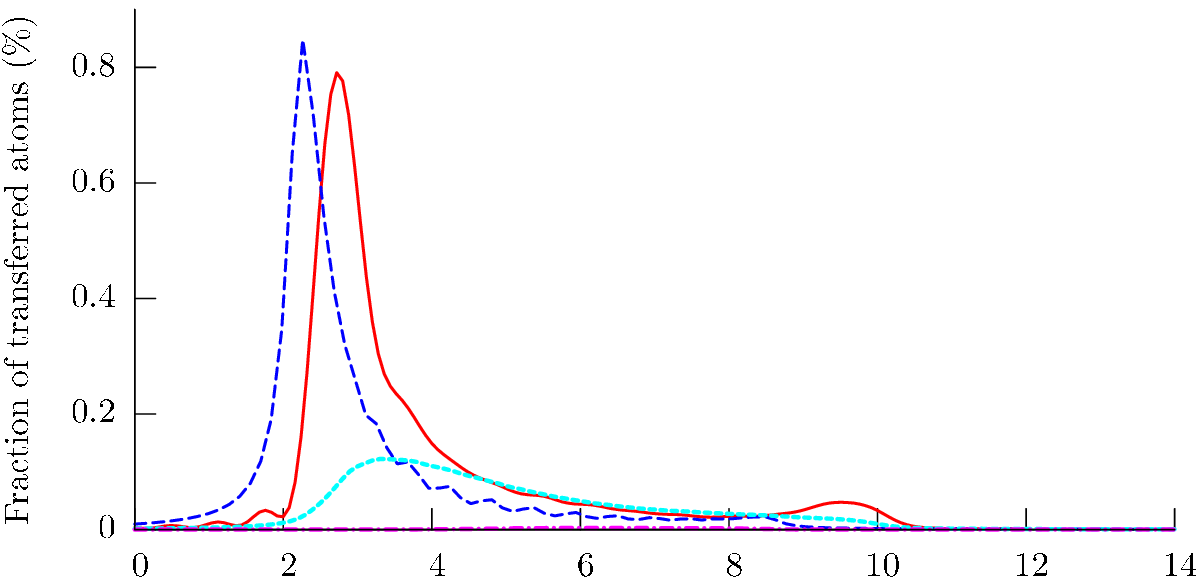}\\
  \includegraphics[width=0.60\textwidth]{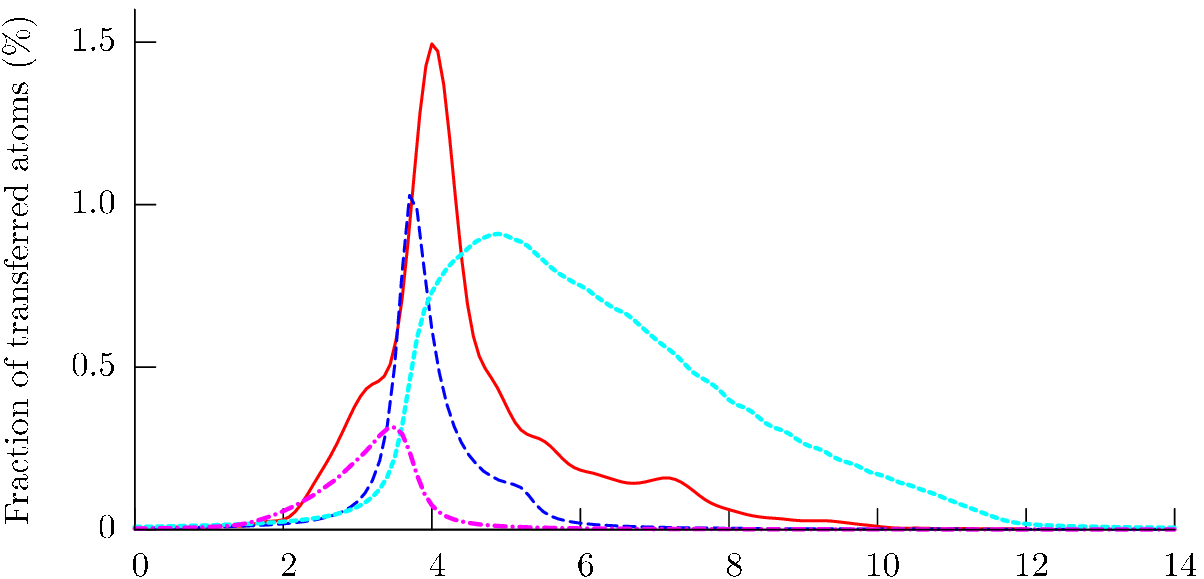}
  \caption{Exact radio-frequency spectra in one-dimensional lattice. 
The time-evolved block decimation method is used for 
calculating the exact
rf-spectra for filling fractions (per spin state) a) 0.3 ($1.1\,J$, $-1.45\,J$), b) 0.5 ($1.1\,J$, $0.0\,J$), and c) 0.7 ($1.1\,J$, $1.45\,J$), with corresponding BCS fitting parameters
shown as ($\Delta$, $\mu$), respectively. The on-site interaction
strength is $U = -5\,J$, rf-coupling strength $\Omega = 0.025\,J$ and
the length of the pulse is $T = 10\,\hbar/J$. Shown are also linear response
(mean-field) spectra and spectra from the RSA model for both scatterings
to higher and lower kinetic energy states (see text for discussion). The RSA 
transition probabilities are not properly normalized and hence the
RSA spectra have been divided by $100$ in order to better
fit them in the picture. We have used 40 lattice sites in these 
simulations.}
  \label{fig:TEBDspectra}
\end{figure}
We use the TEBD algorithm for solving the ground state of the Fermi-Hubbard
Hamiltonian in the absence of the radio-frequency field $\Omega = 0$ and
for empty hyperfine state $|3\rangle$. Starting from this ground state,
we switch on the radio-frequency field and calculate the time evolution.
Fig.~\ref{fig:TEBDspectra} shows the fraction of atoms in state $|3\rangle$
after the radio-frequency pulse as a function of rf-detuning $\delta$.

The BCS theory is known to fail in one-dimensional lattices. Even though the 
mean-field model does reproduce exact ground state energies faithfully,
the behaviour of the pairing gap is phenomenally bad. In order to fix this,
we do not solve the pairing gap $\Delta$ and the chemical potential $\mu$
from the BCS theory but rather derive them from the TEBD ground state
following procedure in~\cite{Marsiglio1997a}. We use these values $\Delta_\mathrm{TEBD}$ and $\mu_\mathrm{TEBD}$ in the
BCS-type quasiparticle dispersion energies 
$E_k^\mathrm{fit} = \sqrt{(\epsilon_k-\mu_\mathrm{TEBD})^2 + \Delta_\mathrm{TEBD}^2}$,
where the lattice dispersion $\epsilon_k = -2J \cos k$ for $k \in [-\pi,\pi]$,
and in the Bogoliubov coefficients for the occupation numbers 
$v_k^2 = \frac{1}{2} \left( 1 - \frac{\epsilon_k - \mu_\mathrm{TEBD}}{E_k}\right)$. 
The linear response rf-spectra, obtained using these TEBD derived parameters
in the quasiparticle dispersion $E_k^\mathrm{fit}$ agree very well with parts
of the observed TEBD spectra. However, the high energy tails of the spectra 
are missing from the linear response spectra, as the spectra are consistently
too narrow. This shows that the BCS-type hole spectral function 
does not capture all of the relevant physics of the rf-photon absorption. 
The question is now whether this is a problem with the BCS-like description
of the ground state (that is, how well does the BCS-like quasiparticle 
dispersion work even when used with the TEBD derived parameters) or with 
the dynamics described by the linear response theory. With this caveat
in mind, let us consider how would the RSA process affect the picture.

Fig.~\ref{fig:TEBDspectra} shows also the RSA spectra from 
the Eqs.~\ref{eq:scatspec3} and~\ref{eq:scatspec}.
This contribution has been separated into two parts,
one corresponding to atom-atom scatterings to higher kinetic energy states 
$\Delta K_{kpq} = E_{k+q}^2 + \epsilon_p^1 - E_k^2 - \epsilon_{p-q}^1 > 0$ 
and one for scatterings to lower kinetic energy states 
$\Delta K_{kpq} < 0$. While the former 
process can be understood as a phonon excitation, the latter process
may be unphysical and possibly an artifact from our approximative approach.
We will neglect these scatterings to lower energy states for a moment and
discuss them later on. 

The combination of the standard linear response spectrum
and the RSA spectrum seems to describe rather well all
qualitative features of the exact TEBD spectrum: Eq.~\ref{eq:linear}
reproduces the main peak of the exact spectrum and Eq.~\ref{eq:scatspec3}
provides the correct width for the spectrum (neglecting the $\Delta K < 0$
contribution). Indeed, the $8\,J$
width of the spectrum can be observed in all TEBD spectra at different
filling fractions and different interaction strengths (although resolution
becomes difficult for weak interactions $-U < 4\,J$). This width
in the one-dimensional lattice is also a general property of 
Eqs.~\eqref{eq:scatspec3} and~\eqref{eq:scatspec}. The low energy end of the
spectrum is provided by low momentum $|2,k=0\rangle$-atoms that scatter from
$|1,p=k_\mathrm{F}\rangle$-atoms at the Fermi surface but the change in
momentum is small ($q=0$). Effectively it corresponds to no scattering at all,
and the resonant energy for the transition is 
$\delta_\mathrm{low} = \epsilon_0^3 - \mu + E_0^2$.
The high energy end of the spectrum is provided by $|2,k = \pi\rangle$-atoms 
at the edge of the lattice band scattering from $|1,p=0\rangle$-atoms in the
bottom of the Fermi sea with momentum exchange $q = -\pi$, yielding
resonant energy $\delta_\mathrm{high} = \epsilon_\pi^3 + E_0^2 + \epsilon_\pi^1 - \epsilon_0^1-\mu$. Thus the total 
width of the spectrum equals $\delta_\mathrm{high} - \delta_\mathrm{low} = 
\epsilon_\pi^3 + \epsilon_\pi^1 - \epsilon_0^1 - \epsilon_0^3 = 8\,J$,
independent of the chemical potential $\mu$ (and hence filling fraction) 
and the excitation gap $\Delta$. This is in 
\emph{excellent, quantitative agreement} with 
the exact TEBD spectra. Notice also that this implies that the width of the
spectrum is actually insensitive to the way how the $\Delta$ and $\mu$ are
determined, and even the values from the BCS theory would yield the same
total width; only the positions of the peaks would be affected. And finally,
the quasiparticle dispersions vanish from the calculation of the total width,
and hence even if one would relax the assumption of the BCS-type quasiparticle
dispersion $E_k^\mathrm{fit}$ somewhat, the result would remain the same.

At high filling fractions our model predicts a low energy tail extending
beyond the $8\,J$ width in disagreement with the TEBD spectra. The tail
is caused by scatterings to lower energy states $\Delta K_{kpq} < 0$. Since the
$|1\rangle$-atoms are forbidden from scattering to lower kinetic energy states
due to the presence of a well-formed Fermi sphere,
the lowering of the energy is possible only if the \emph{increase} in the 
kinetic 
energy of the $|1\rangle$-atom is lower than the \emph{decrease} in the 
quasiparticle energy of the $|2\rangle$-atom. Thus we have condition
\begin{equation}
   0 > \Delta K_{kpq} = (E_{k+q} - E_k) + (\epsilon_{p-q} - \epsilon_p).
\end{equation}
In the long scattering wavelength limit $q\rightarrow 0$ this yields
\begin{equation}
   \frac{d E_k}{d k} < -\frac{d \epsilon_p}{dp}|_{k = k_\mathrm{F}},
\label{eq:diffK}
\end{equation}
where the right-hand side must be evaluated close to the Fermi surface because
of the well-formed Fermi sphere for $|1\rangle$-atoms. The left hand
side yields $\frac{\epsilon_k - \mu}{E_k} \frac{d\epsilon_k}{dk}$, 
where the prefactor $(\epsilon_k-\mu)/E_k$ has magnitude less than $1$.
However, since the quasiparticles can populate any momentum state $k$, it
is possible that Eq.~\eqref{eq:diffK} is satisfied for some $k$ as long
as the single particle dispersion is not too rapidly increasing function
at the Fermi momentum $k_\mathrm{F}$. In particular, at high fillings
the single-particle dispersion $\epsilon_k$ becomes concave at the Fermi 
surface 
and the scatterings to lower kinetic energy states $\Delta K < 0$ 
becomes possible. At half filling, the single-particle
dispersion has the highest slope and no $\Delta K < 0$ scatterings are
possible. At low fillings there are some scattering channels as well but
the contribution is very small as compared to the scatterings to higher
energy states $\Delta K > 0$ (typically at least one order of magnitude
lower).

\section{Discussion}

In conclusion, we have formulated a first order scattering theory which 
adds to the understanding of interdisciplinary spectroscopies.
Despite being of the first order in the photon coupling, the theory
goes beyond standard mean-field linear response theories
by incorporating an approximative coupling to phonon-like modes.
Similar phenomena have been extensively studied in nuclear physics since
1960's~\cite{Soloviev}, and it would be interesting to compare the present
approach with the models used in nuclear physics.
The resonant scattering process in ultracold atom gases becomes increasingly
important with stronger interactions and its effect can be seen in a 
wide range of experiments -- here we have analyzed seven of them.

{\ack
This work was supported by the National Graduate School in Materials Physics
and Academy of Finland (Project No. 217043). We acknowledge inspiring 
discussions with F. Massel and A.-P. Jauho, and the use of CSC - IT Center 
for Science Ltd computing resources.}

\section*{References} 
\bibliographystyle{unsrt}
\bibliography{paperi.bib}
\end{document}